\documentclass[aps,showpacs,twocolumn,longbibliography,nofootinbib,secnumarabic,eqsecnum,superscriptaddress]{revtex4-1}

\usepackage{graphicx}
\usepackage{dcolumn}
\usepackage{bm}
\usepackage{braket}
\usepackage{leftidx}
\usepackage{cancel}
\usepackage{amsmath}
\usepackage{epstopdf}
\usepackage{color}
\usepackage[mathcal]{eucal}
\usepackage{float}

\usepackage{yfonts}
\usepackage{color}

\usepackage{url}

\usepackage{scalerel}

\usepackage{calligra}

\usepackage{pbsi}
\usepackage[T1]{fontenc}

\begin{document}
\date{\today}
\title{Quantifications for multi-mode entanglement}

\author{Mehmet Emre Tasgin}
\email{metasgin@hacettepe.edu.tr}
\affiliation{Institute of Nuclear Sciences, Hacettepe University, 06800, Ankara, Turkey}


\author{M. Suhail Zubairy}
\affiliation{Institute of Quantum Studies and Department of Physics, Texas A {\&} M University, College Station, TX
77 843-4242, USA}

\begin{abstract}
We introduce two independent quantifications for 3-mode and 4-mode entanglement. We investigate the conversion of one type of nonclassicality, i.e. single-mode nonclassicality, into another type of nonclassicality, i.e. multi-mode entanglement, in beam-splitters. We observe parallel behavior of the two quantifications. The methods can be generalized to the quantification of any multi-mode entanglement. 
\end{abstract}

\maketitle

Multi-mode continuous variable entanglement is a key element for quantum information processing. Quantum teleportation among multiple parties~\cite{BookQuantumTeleportation&Entanglement2011}, quantum computation with clusters~\cite{menicucci2006universal,raussendorf2001one}, e.g. quantum networks,~\cite{armstrong2012programmable} and quantum internet~\cite{kimble2008quantumInternet}, all, necessitate the presence of multi-mode entanglement. Multi-mode entanglement can be generated using several  number of beam-splitters which converts the single-mode nonclassicality~(SMNc), e.g. quadrature squeezing, of input beam(s) into multi-mode entanglement at the outputs of the beam splitters~\cite{Loock&Braunstein2000multipartite}. Hence, mechanism of the conversion of nonclassicality into multi-mode entanglement, via beam splitters, is important for distribution of the entanglement among multi-parties used in quantum information.

One can notice that quantum nonclassicality is like "energy". That is, it can be converted into different forms. For instance, single-mode nonclassicality~(SMNc), two-mode entanglement~(TME) and many-particle entanglement~(MPE) are different forms nonclassicality~(Nc) and can be converted to each other via linear interactions.~\footnote{Also the criteria witnessing them can be transformed into each other~\cite{tasgin2017many}.}
A quadrature-squeezed light~(SMNc) can transfer its squeezing to an ensemble of atoms~\cite{PolzikPRL1999spinsqz,TothPRA2018SpinSqz,vidal2006concurrence} as spin-squeezing~\cite{kitagawa1993squeezed}. This creates MPE~\cite{sorensen2001Nature,tasgin2017many} in the ensemble.  Similarly, interaction of an ensemble with two entangled beams~(TME) also creates MPE in the ensemble~\cite{regula2018converting,Tasgin&MeystrePRA2011}. It is also possible to convert the nonclassicality of a single-mode, e.g. quadrature-squeezing, into two-mode entanglement at the output of a beam-splitter~(BS)~\cite{Kim:02,Asboth:05}. Generation of TME in a BS necessitates a nonclassical input light~\cite{Kim:02}. 

Recent works~\cite{ge2015conservation,arkhipov2016nonclassicality,arkhipov2016interplay,vcernoch2018experimental} show that there appears a conservation-like relation between the generated TME and the remaining SMNc in a BS. A BS cannot convert all of the nonclassicality of the input beam into TME at the output. There still remains some SMNc in the two output modes~\cite{ge2015conservation}. However, the total SMNc decreases with respect to the input one. The form of the analytical expressions for the maximum TME extracted at the BS output, $E_{\cal N}={\rm max}\{0,-\frac{1}{2}\log_2(\lambda_{1,{\rm sm}}^{\rm (in)} \lambda_{2,{\rm sm}}^{\rm (in)}) \}$~\cite{Tahira:09,Li:06}, suggests us to quantify the SMNc in terms of a noise-area~\cite{ge2015conservation}, $\Omega=\lambda_{1,{\rm sm}} \lambda_{2,{\rm sm}}$. Here, $\lambda_{i,{\rm sm}}$ is the minimum noise of the $i$th beam input into the BS, where $\lambda_{i,{\rm sm}}<1$ implies the presence of squeezing in the $i$th beam. For mixing with vacuum or a coherent state, one of the modes becomes $\lambda_{1,{\rm sm}}=1$.

As an illuminating example: if a squeezed beam is mixed with a thermal noise~\cite{Tahira:09} at the input channels of a BS, the output modes are entangled only if the noise-area $\Omega=\lambda_{\rm sqz}\:(1+2\bar{n})<1$, where $\lambda_{\rm sqz}<1$ is the reduced noise of the squeezed beam and $\bar{n}$ is the mean number of photons in the thermal noise. Also Ref.~\cite{Li:06} shows that the maximum amount of entanglement extractable at the BS output is $E_{\cal N}=-\frac{1}{2}\log_2(2\nu_-)=-\frac{1}{2}\log_2(\lambda_{1,{\rm sm}}^{\rm (in)} \lambda_{2,{\rm sm}}^{\rm (in)}) \}$ if any two Gaussian states are mixed in the BS input. Here, one can observe that the smallest symplectic eigenvalue of an inseparable system~\cite{adesso2004extremal,Vidal:02}  becomes $\mu=2\nu_-=\lambda_{1,{\rm sm}}^{\rm (in)} \lambda_{2,{\rm sm}}^{\rm (in)}$, which is actually the input noise-area. In Ref.~\cite{ge2015conservation}, we further show that the output TME is proportional to the change~(increase) in the noise-area $S_N=\log_2(\lambda_{1,{\rm sm}}^{\rm (out)} \lambda_{2,{\rm sm}}^{\rm (out)}) - \log_2(\lambda_{1,{\rm sm}}^{\rm (in)} \lambda_{2,{\rm sm}}^{\rm (in)})$ of the out beams with respect to the input beams. A geometric demonstration of this SMNc$\to$TME swap can be found in Fig.~1 of Ref.~\cite{tasgin2019anatomy}. More interestingly, well-known TME criteria, like Duan-Giedke-Cirac-Zoller~\cite{DGCZ_PRL2000} and Hillery-Zubairy~\cite{Hillery&ZubairyPRL2006}, actually do search for a noise-area below unity in BS-like rotations~\cite{tasgin2019anatomy}.

\begin{figure}
\includegraphics [width=0.27\textwidth]{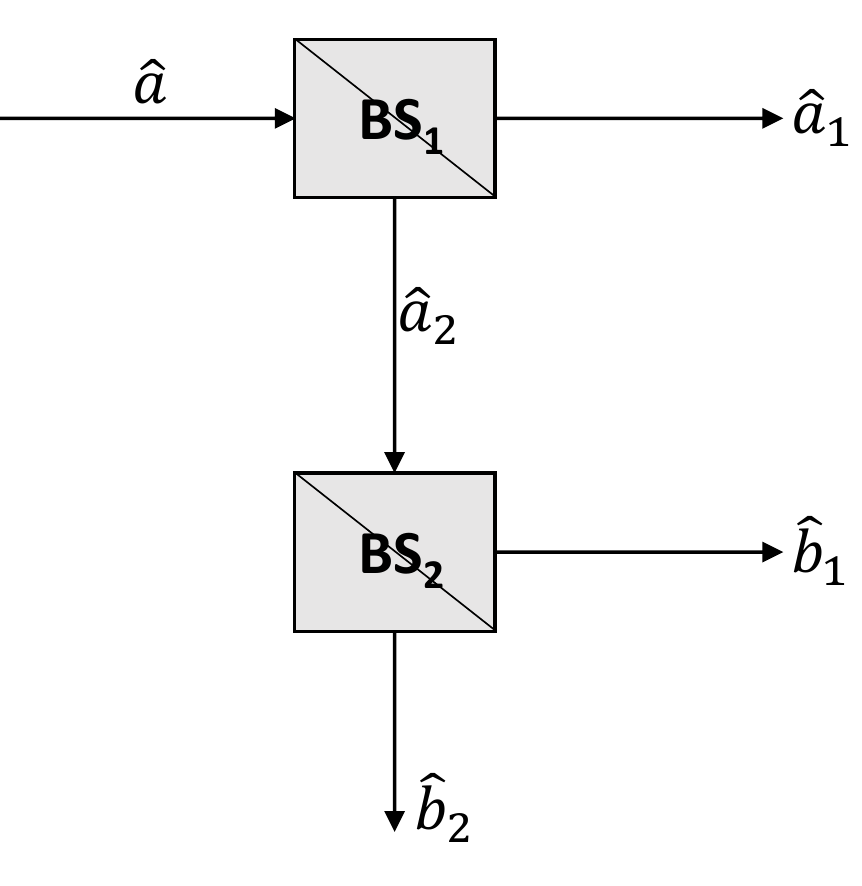} 
\caption{Creation of 3-mode entanglement~(3ME) using 2 beam-splitters. We examine the 3ME of $\hat{a}_1$, $\hat{b}_1$, $\hat{b}_2$ modes. We also investigate the conversion of the single-mode nonclassicality into 3ME, see Fig.~\ref{fig2}.}
\label{fig1}
\end{figure}

In the present work, we investigate the swapping of single-mode nonclassicality into multi-mode entanglement~(MME). In particular, we study three-mode entanglement~(3ME) and four-mode entanglement~(4ME) after two~(Fig.~\ref{fig1}) and three beam-splitters~(Fig.~\ref{fig3}), respectively. We define the remaining SM nonclassicalities in the 3 modes (or 4 modes) as a noise-volume (or a 4D noise-volume), $\Omega=\lambda_{1,{\rm sm}}^{\rm (out)}\:\lambda_{2,{\rm sm}}^{\rm (out)} \: \lambda_{3,{\rm sm}}^{\rm (out)}$. Here, $\tau_i^{\rm (out)}=1-\lambda_{i,{\rm sm}}^{\rm (out)}$ is the degree of the remaining SMNc, i.e. the nonclassical depth~\cite{lee1991measure}, of each mode after the BSs. It is calculated by wiping out the entanglement (correlations) between the modes~\cite{ge2015conservation}. We show that the SMNc decreases (noise-volume $\Omega$ increases) after the BSs while the the 3ME increases.

In the quantification of 3ME we use two different (independent) approaches. ({\it i}) We multiply the three noises associated with the symplectic eigenvalues ($\nu_1$,$\nu_2$,$\nu_3$) of the partial transposed system. $\nu_1$, for instance, is the smallest symplectic eigenvalue of the 3-mode system when the 1st mode is partial transposed~\cite{adesso2004extremal,Vidal:02}. That is, smaller values of the eigenvalue compared to unity, i.e. $2\nu_1<1$, imply stronger entanglement of the 1st mode with the system composed of (2nd + 3rd) modes~\cite{plenio2005logarithmic}. Similarly, $2\nu_2<1$ refers to the inseparability of the 2nd mode from the system of (1st + 3rd) modes. Hence, $\mu=2\nu_1\:2\nu_2\:2\nu_3$ refers to a kind of 3-mode entanglement strength, where $2\nu_i>1$ implies the absence of a genuine 3-mode entanglement.

({\it ii}) Second, we use an alternative method given in Sec.~II.3.2 of Ref.~\cite{tasgin2019anatomy}. The idea is very simple: Nonclassical depth $\tau$ quantifies the whole nonclassicality, i.e. SMNc + entanglement, in a multi-mode system~\cite{Li:06}. If we remove the unconverted (unused) SM nonclassicalities (after the BSs) from the noise-matrix, then the remaining nonclassicality is due to the entanglement only~\footnote{In Ref.~\cite{ge2015conservation}, we do the reverse. We remove the correlations in the noise matrix and examine the remaining SM nonclassicalitites.}. Nonclassicality of a single-mode state can be determined by introducing a Gaussian filter function transformation on the Glauber-Sudarshan $P$-function, i.e. $P(\alpha,\tau)= \int d^2\alpha' \: \exp\{ -|\alpha-\alpha'|^2/\tau \} \: P(\alpha')\:/\tau\pi$. So that, the new $P$-function is non-negative~\cite{lee1991measure}. This corresponds to injecting noise which destroys the nonclassicality~\cite{vogelPRA2010Filters}. A similar method can be used also to determine the nonclassicality of a multi-mode system~\cite{Li:06}. Here, in difference to Ref.~\cite{Li:06}, we introduce different $\tau$'s for each mode. This way, we prevent the injection of unnecessary noise ($\tau$) by constraining $\tau_1=\tau_2=\tau_3=\tau$.

We emphasize that, our aim, in this short manuscript, is to introduce the basics of  two possible quantifications for the multi-mode entanglement~(MME). We do not aim to present a detailed analysis on MME, but we suffice with demonstrating that SMNc, quantified as a noise-volume, is converted into MME.

\vspace{0.5 cm}
{\bf 3-mode entanglement}
\vspace{0.15 cm}

We study the system depicted in Fig.~\ref{fig1}. A single-mode~(SM) nonclassical state $\hat{a}$ is mixed with vacuum noise in a BS with two output states $\hat{a}_1$ and $\hat{a}_2$. One of the output modes, $\hat{a}_2$, is input to a second BS, mixed with vacuum, resulting two new output modes $\hat{b}_1$ and $\hat{b}_2$. We examine the 3-mode entanglement~(3ME) of $\hat{a}_1$, $\hat{b}_1$, and $\hat{b}_2$ modes.

In Fig.~{\ref{fig2}a and Fig.~\ref{fig2}b, we examine the 3ME of the $\hat{a}_1$, $\hat{b}_1$, $\hat{b}_2$ modes using the two methods, ({\it i}) and ({\it ii}), respectively. We observe a similar behavior for the two quantifications. In Fig.~\ref{fig2}c, one can observe that the SM nonclassicalities remaining in the $\hat{a}_1$, $\hat{b}_1$, $\hat{b}_2$ modes decreases (noise-volume increases), while the 3ME increases. In all Fig.~\ref{fig2}a-\ref{fig2}c, smaller $\mu$, $\eta$ and $\cal N$ imply stronger 3ME or SMNc. The first BS fed with a nonclassical light of squeezing parameter~\cite{ScullyZubairyBook} $r=0.1$. We fix the angle of the first BS to $\theta_{{}_{\rm BS_1}}=\pi/4$ and vary the angle of the second BS $\theta_{{}_{\rm BS_2}}$.A similar behavior is obtained for varying $\theta_{{}_{\rm BS_1}}$ with a fixed $\theta_{{}_{\rm BS_2}}$.
\begin{figure}
\includegraphics [width=0.5\textwidth]{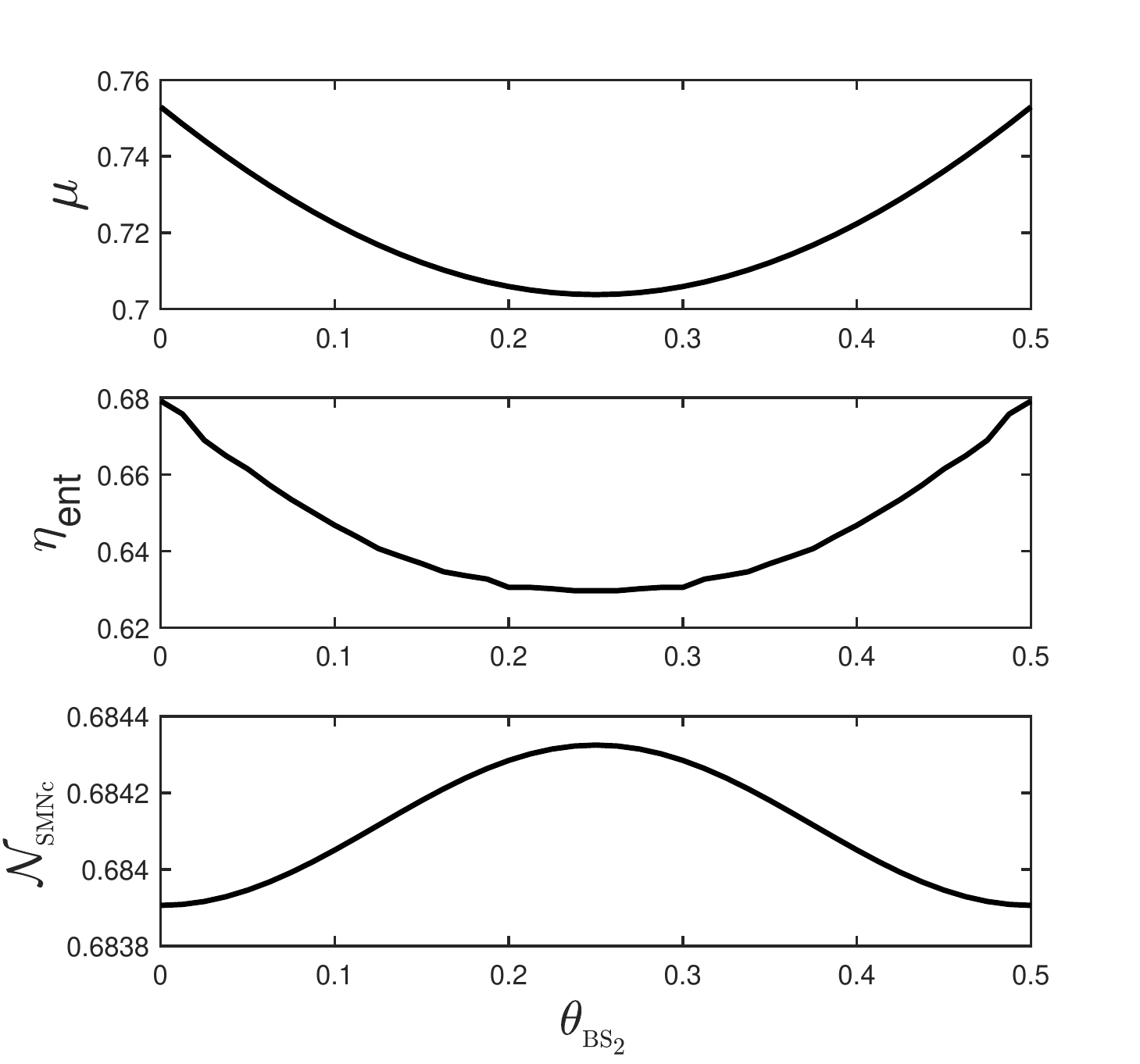} 
\caption{Conversion of SMNc into 3-mode entanglement~(3ME). (a) Quantification of 3ME via symplectic eigenvalues $\mu=2\nu_1\:2\nu_2\:2\nu_3$ where $2\nu_i$ is obtained by the partial transposition of the $i$th mode. See method ({\it i}) in the text. (b) Quantification of the entanglement by wiping out the single-mode nonclassicalities from the noise-matrix, where the remaining nonclassicality refers to the entanglement only. See method ({\it ii}) in the text. (c) The remaining single-mode nonclassicality~(SMNc) in the output modes when the correlations are wiped out, in stead of the SMNc. Smaller $\mu$,$\eta_{\rm ent}$,${\cal N}<1$ implies a stronger entanglement/nonclassicality strength.
  }
\label{fig2}
\end{figure}

$\mu$, $\eta_{\rm ent}$ and $\cal N$ are calculated from the 3 mode noise matrix as follows. 

{\small{\bf Calculation of}} {\boldmath$\mu$}.--- We compose the 6$\times$6 noise-matrix $V_{ij}=\langle\hat{\xi}_i\hat{\xi}_j+\hat{\xi}_j\hat{\xi}_i\rangle/2 - \langle\hat{\xi}_i\rangle\langle\hat{\xi}_j\rangle$  for the 3 mode system in the real representation~\cite{simon1994quantum} by introducing the operator $\hat{\xi}=[\hat{x}_1\:,\:\hat{p}_1\:,\:\hat{x}_2\:,\:\hat{p}_2\:,\:\hat{x}_3\:,\:\hat{p}_3]$, where $\hat{x}_i=(\hat{c}_i^\dagger+\hat{c}_i)/\sqrt{2}$ and $\hat{p}_i=i(\hat{c}_i^\dagger-\hat{c}_i)/\sqrt{2}$ are obtained using $\hat{c}_i=\hat{a}_1$,$\hat{b}_1$,$\hat{b}_2$, respectively. Partial transpose~(PT) operation, e.g., on the 1st mode is equivalent to $\hat{p}_1\to-\hat{p}_1$ in the noise-matrix~\cite{SimonPRL2000}. If the 1st mode is separable from the system of (2+3) modes, symplectic eigenvalues of the partial transposed noise-matrix must be all larger than $1/2$~\cite{Vidal:02,adesso2004extremal}. (So, smallest eigenvalue $\nu_1$ also satisfies $2\nu_1>1$.) The symplectic eigenvalues of the partial transposed noise-matrix can alternatively be calculated as ${\rm eig}[i\chi_1V]$ where $\chi_1$ is the 6$\times$6 matrix $\chi_1=[-J\:,\: 0_{2\times2}\:,\:0_{2\times2}\: ; \: 0_{2\times2}\:,\:J\:,\:0_{2\times2} \:; \: 0_{2\times2}\:,\:0_{2\times2},\:J]$ with $0_{2\times2}$ is 2$\times$2 matrix of zeros and $J=[0\:,\:1\: ; \:-1\:,0]$. For $\chi_2$ and $\chi_3$ the "-" sing must be in the second and third $J$, respectively. ${\rm eig[i\chi_1 V]}$ yields only a single eigenvalue with $2\nu_1<1$. Similarly, $\nu_2$ ($\nu_3$) is the only $2\nu_2<1$ ($2\nu_3<1$) eigenvalue from the partial transposition of the 2nd (3rd) mode.

We remind one more time that introduction of the noise-volume $\mu=2\nu_1\:2\nu_2\:2\nu_3$ via the symplectic eigenvalues follows from the observation $\log_22\nu_-=\log_2(\lambda_{1,{\rm sm}}^{\rm (in)}\lambda_{2,{\rm sm}}^{\rm (in)})$ for a single BS~\cite{Tahira:09,Li:06,ge2015conservation}, where $2\nu_-=\lambda_{1,{\rm sm}}^{\rm (in)}\lambda_{2,{\rm sm}}^{\rm (in)}$ determines the maximum entanglement extractable from the input noise-area $\lambda_{1,{\rm sm}}^{\rm (in)}\lambda_{2,{\rm sm}}^{\rm (in)}$.

{\small{\bf Calculation of}} {\boldmath$\eta_{\rm ent}$}.--- The nonclassical depth, associated only with the entanglement, is calculated as follows. We first transform the real noise-matrix $V$ into the complex representation~\cite{simon1994quantum} $V^{\rm (c)}={\cal C}V{\cal C}^\dagger$, where ${\cal C}=[{\cal C}_1,0_{2\times2}, 0_{2\times2};  0_{2\times2}, {\cal C}_1, 0_{2\times2}; {\cal C}_1, 0_{2\times2}, 0_{2\times2}]$ with ${\cal C}_1=[1,i; 1,-i]/\sqrt{2}$. In the 6$\times$6 complex noise-matrix $V^{\rm (c)}$, we wipe out the SM nonclassicalities of the 3 modes, e.g. in the 1st mode, by replacing the $[V_{11},V_{12}; V_{2,1},V_{2,2}]$ with $[1/2,0;0,1/2]$, the noise-matrix for vacuum or a coherent state~\cite{Tahira:09}. We wipe out the SMNc of the other 2$\times$2 block-diagonals similarly. Then, we obtain the new noise-matrix $V_{\rm ent}^{\rm (c)}$, where the nonclassical depths $\tau_{1,2,3}$ accounts the entanglement only. For Gaussian states, we consider here, this can be performed by calculating the $\tau_1$, $\tau_2$, $\tau_3$ which makes all ${\rm eig[V_{ent}^{\rm (c)} + \text{\boldmath $\tau$}]}$ positive where {\boldmath $\tau$}$={\rm diag[\tau_1,\tau_1,\tau_2,\tau_2,\tau_3,\tau_3]}$. Ref.~\cite{Li:06} assigns a single $\tau_1=\tau_2=\tau_3=\tau$ for the {\boldmath $\tau$} matrix which certainly increases the injected noise. We quantify the nonclassicality of $V_{\rm ent}^{\rm (c)}$, or the entanglement, by choosing the minimum of $\tau_{\rm ent}=[\tau_1\tau_2\tau_3]_{\rm min}$, or $\eta_{\rm ent}=\left[(1-2\tau_1)(1-2\tau_2)(1-2\tau_3)  \right]_{\rm max}$ in terms in terms of the injected noise-volume. We note that, for a single-mode, $\lambda_{\rm sm}=(1-2\tau)$ corresponds to the reduced noise of that particular mode. 

{\small{\bf Calculation of}} {\boldmath$\cal N_{{}_{\rm SMNc}}$}.--- In the calculation of SMNc $\cal N_{{}_{\rm SMNc}}$, this time, we wipe out the correlations in the noise-matrix and left with the three 2$\times$2 block-diagonals. 2$\times$2 matrices give the SMNc associated with each mode~\cite{ge2015conservation}. Then, we introduce the SMNc noise-volume ${\cal N}_{{}_{\rm SMNc}}=(1-2\tau^{{}^{\rm (SMNc)}}_1)(1-2\tau^{{}^{\rm (SMNc)}}_2)(1-2\tau^{{}^{\rm (SMNc)}}_3)$.

\vspace{0.5 cm}
{\bf 4-mode entanglement}
\vspace{0.15 cm}
\begin{figure}[htb!]
\includegraphics [width=0.47\textwidth]{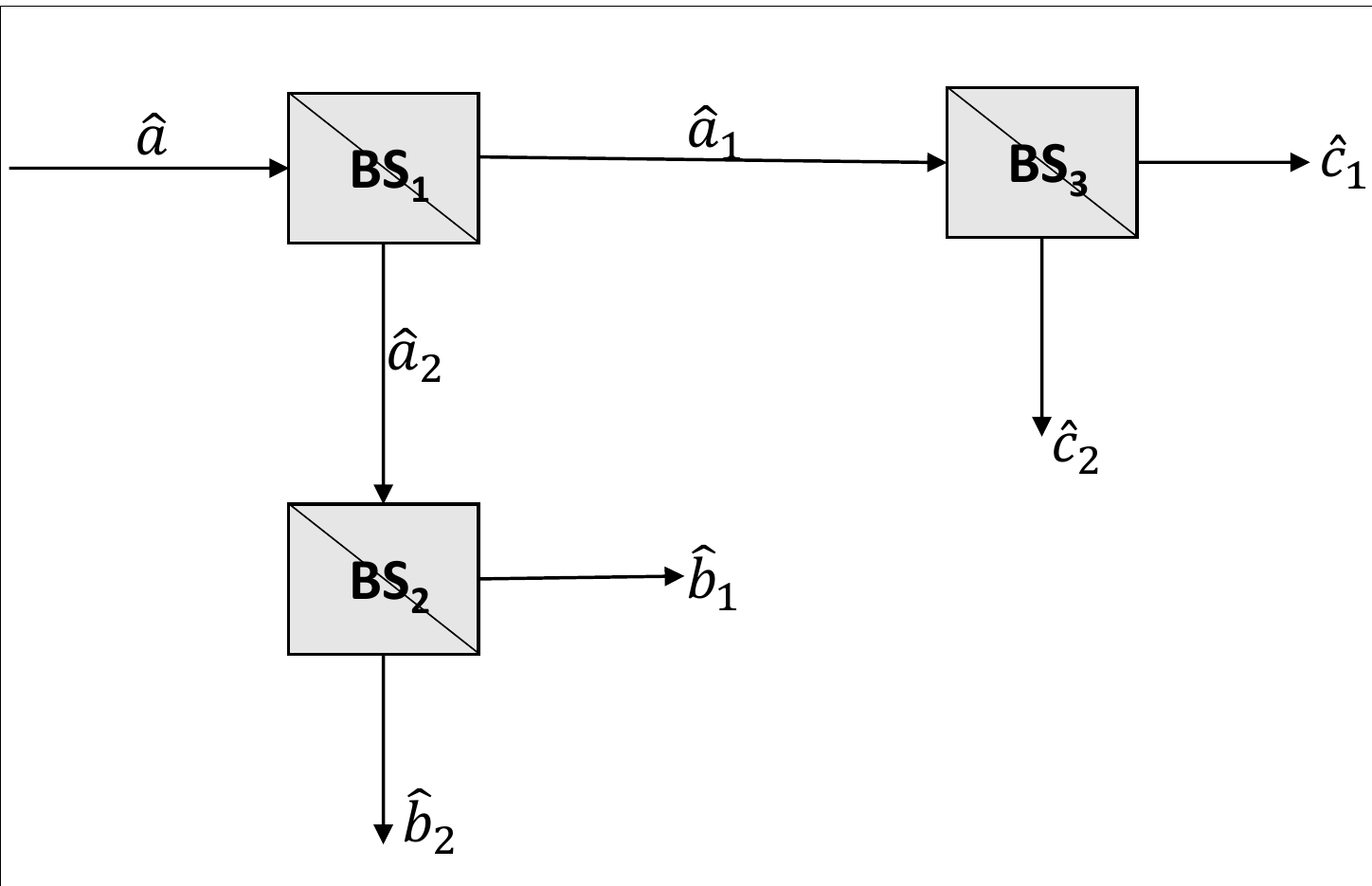}
\caption{Creation of 4-mode entanglement~(4ME) with 3 beam-splitters. We examine the 4ME of $\hat{b}_1$, $\hat{b}_2$, $\hat{c}_1$, $\hat{c}_2$ modes. We also investigate the conversion of the single-mode nonclassicality into 4ME, see Fig.~\ref{fig4}.}
\label{fig3}
\end{figure}

We perform similar calculations also for a 4-mode system given in Fig.~\ref{fig3}. We obtain the same behavior depicted in Fig.~\ref{fig4}.
\begin{figure}
\includegraphics [width=0.5\textwidth]{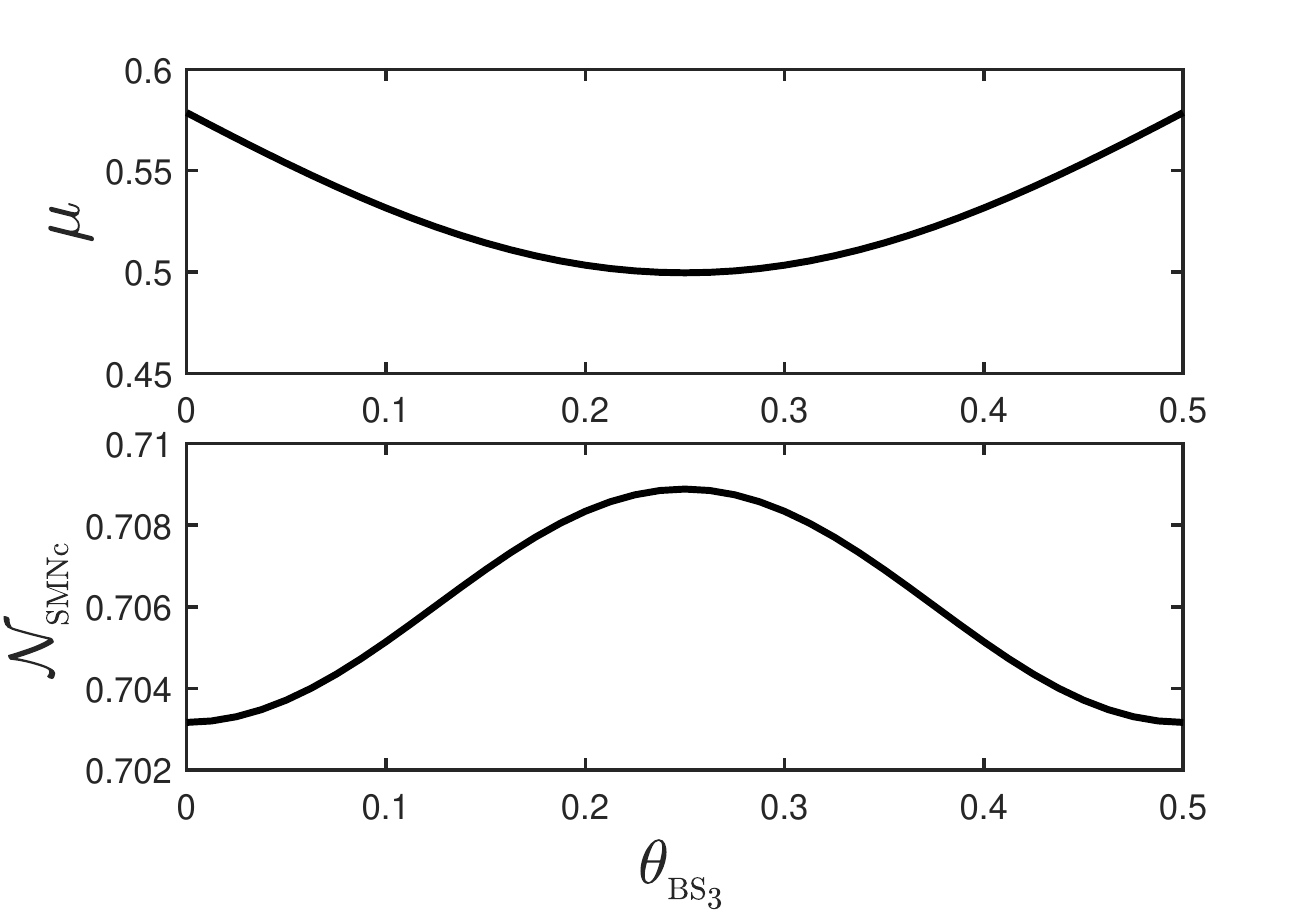} 
\caption{The same of Fig.~\ref{fig2} for the 4-mode entanglement~(4ME) for the system given in Fig.~\ref{fig3}. We observe that SMNc is swapped into 4ME. We cannot plot $\eta_{\rm ent}$ since its calculation for a 4D system is seriously time-consuming. 
  }
\label{fig4}
\end{figure}

\vspace{0.5 cm}
{\bf Summary}
\vspace{0.15 cm}

In summary, we introduce quantifications for 3-mode and 4-mode entanglement via two independent methods. We demonstrate how single-mode nonclasscality is converted into 3-mode and 4-mode entanglement. We quantify all nonclassicalities in terms of noise-volume. A smaller noise-volume implies a stronger nonclassicality or entanglement. The method we introduce here can be generalized to other multi-mode entanglement which has fundamental importance in quantum communication.

\vspace{0.5 cm}
{\bf Acknowledgments}
\vspace{0.15 cm}

MET acknowledges support from TUBITAK Grant No: 1001-117F118, TUBA-GEBIP Award 2017, Hacettepe University BAP Grant No: FBI-2018-17423.

\bibliography{bibliography}
  
\end{document}